\documentclass[12pt]{article}

\newcommand{\Break}{ \right. \nonumber \\ &{}& \left. }
\renewcommand{\arraystretch}{1.5}

\newcommand{\ice}[1]{\relax}

\newcommand{\beq}{\begin{equation}}
\newcommand{\eeq}{\end{equation}}
\newcommand{\bea}{\begin{eqnarray}}
\newcommand{\eea}{\end{eqnarray}}

\newcommand{\api}{\frac{\alpha_s}{\pi}}

\newcommand{\ba}{\begin{array}} 
\newcommand{\ea}{\end{array}} 
\newcommand{\ds}{\displaystyle} 
\newcommand{\as}{\alpha_s}

\newcommand{\msbar}{\overline{\mbox{MS}}}

\newcommand{\dsp}{\displaystyle}

\newcommand{\EQN}{\label}

\newcommand{\re}[1]{(\ref{#1})}

\begin{document}

\begin{titlepage}
\noindent
%
%
\hfill TTP98--18\\
\mbox{}
\hfill  May 1998  \\   
\mbox{}
\hfill hep-ph/9805335\\
\hfill {}

\vspace{0.5cm}
\begin{center}
  \begin{Large}
  \begin{bf}
Determining 
the Strange Quark Mass in Cabibbo Suppressed Tau Lepton Decays
  \end{bf}
  \end{Large}
%
%
  \vspace{0.8cm}

  \begin{large}
K.G.Chetyrkin\footnote{On leave from Institute for Nuclear Research
of the Russian Academy of Sciences, Moscow, 117312, Russia.},
J.H.~K\"uhn  \\[3mm]
    Institut f\"ur Theoretische Teilchenphysik\\
    Universit\"at Karlsruhe\\
    Kaiserstr. 12,    Postfach 6980\\[2mm]
    7500 Karlsruhe 1, Germany\\[5mm]
A.A.~Pivovarov$^1$\\
Institut f\"ur Physik,\\
Johannes-Gutenberg-Universit\"at \\
Staudinger Weg 7, D-55099 Mainz, Germany
\end{large}

\vspace{0.8cm}

{\bf Abstract}
\end{center}
\begin{quotation}
\noindent

In this work radiative corrections in the total hadronic decay rate of
the $\tau$ lepton and some moments of its differential distributions
are studied employing perturbative QCD and the operator product
expansion.  We calculate quadratic quark mass corrections in the
strange mass to the decay rate ratio $R_{\tau}$ to the order ${\cal
O}(\alpha_s^3 m^2)$ and find that they contribute appreciably to the
Cabibbo suppressed decay modes of the $\tau$-lepton.  Using the
results of a recent experimental analysis, we obtain 
$m_s(\mbox{1 GeV}) = 200 \pm 40_{exp} \pm 30_{th}$ MeV.
\end{quotation}
\end{titlepage}

\section{Introduction}
\renewcommand{\arraystretch}{2}

With an ever increasing number of $\tau$ leptons observed by the four
LEP experiments and by CLEO the Cabibbo suppressed $\tau$ decays have
become one of the important topics of recent experimental analysis
\cite{exp:data1}. The study of exclusive channels allows to determine
hadronic resonance 
parameters, to test the predictions in the chiral limit and,
quite generally, to explore the hadronic current in the low energy
region.
Multi-differential   angular distributions of mesons   can  be used to
measure   the polarization of  the   $\tau$ lepton and to  furthermore
disentangle the various  spin  parity contributions of hadronic 
states with $J^P  = 0^+$,
$0^-$, $1^-$ or $1^+$ induced by the  (non-) conserved parts of vector
and  axial vector current respectively  \cite{KM}. 
Complementary  studies are
based  on   the inclusive decay  rate   derived  from the semileptonic
branching  ratio  or the $\tau$ life  time
(see e.g. \cite{SchTra84,Bra88,NarPic88,Bra89,BraNarPic92}).  
The  determination of the
strong  coupling constant $\alpha_s$ has been   a focal point in these
investigations. The small reduction of the Cabibbo   suppressed rate
(relative to  the  massless prediction) has been  used recently
\cite{Chen} even to extract a value for the strange quark mass. The 
analysis was based on the total rate and a theoretical calculation
\cite{Chetyrkin93} including quadratic mass terms with coefficients
of order $\as^2$.
Recently, the authors of \cite{Chetyrkin93} have revised their
analysis and corrected an error in the published
numerical 
value of the $\as^2$ term. As a result the contribution from that term
has increased dramatically by a factor of about four 
\cite{Maltman,Chetyrkin93-update}.  This new
development calls for a fresh look at the possibility for an $m_s$
determination from  $\tau$ decays.

In this work the theoretical analysis will be extended and improved in
various ways: 
\begin{itemize}
\item[$\bullet$]
QCD corrections to the mass terms of order
$\alpha_s^3$ will be calculated 
for some of the moments of the spectral functions
which if included lead to a sizeable shift of
the predictions. 
\item[$\bullet$]
It is proposed to use the techniques of \cite{KM} to
separate the states of different spin parity, allowing thus for quark
mass determinations from four independent spectral functions.
Eventually even different moments of their respective spectral
functions might be considered, leading to additional tests of the
method. 
\item[$\bullet$]
The method \cite{tau:resum} of resummation of effects from the running
of both the coupling constant and the strange quark mass along the
contour of integration in the complex plane through the
renormalization group improvement is used for $\tau$ lepton
observables to provide better convergence of the perturbative series.
\end{itemize}
The configurations  with $0^+$ and $0^-$ and 
{ non-vanishing  invariant masses}
are strictly forbidden in the limit of massless quarks.
The integrated rate, i.e. the lowest  moment of the spin separated 
distribution exhibits a remarkable dependence on the perturbative mass
of the strange quark
and a nonperturbative parameter at the same time. Higher moments
are free from  this nonperturbative constant, allowing thus, at least 
in principle, a new determination of the strange quark mass.

The paper is organised as follows. In the next section some general
relations are given and our notation is fixed.  The observables to be
analysed in $\tau$ lepton physics are introduced and the stage is set
for their perturbative analysis. In Sect. 3 some new features are
described which appear due to the mass terms in the Cabibbo suppressed
channel. Explicit expressions for the coefficients of quadratic mass
terms are given.  In the same section both the finite order and
resumed observables are presented. Sect. 4 concentrates on
nonperturbative corrections to spin zero contribution. In sect. 5
numerical results are given. The last section 6 contains our
conclusion about the possibility and accuracy of a strange quark mass
determination from an inclusive analysis of Cabibbo suppressed modes.

\section{Generalities}
In order to set the framework of the subsequent discussion, which
includes quark mass effects and the separation of the spin one and
spin zero contributions, we repeat below a number of  generalities
about the theoretical analysis of $\tau$ lepton observables both in
massless and massive case and also introduce our notations for
necessary quantities which 
can be found in  earlier literature (see, e.g. 
Refs.~\cite{BraNarPic92,phys_report}).

\subsection{Correlators}

Physical $\tau$ lepton observables
are related to correlators of 
vector and axial vector currents of light quarks
that are defined as follows
\begin{equation}
\begin{array}{ll}
\Pi^{V/A}_{\mu\nu,ij}(q,m_i,m_j,m{},\mu,\alpha_s) & =
\displaystyle i \int dx e^{iqx}
\langle
T[\, j^{V/A}_{\mu,ij}(x) (j^{V/A}_{\nu,ij})^{\dagger} (0)\, ] \rangle
\\ &  = \displaystyle 
g_{\mu\nu}  \Pi^{[1]}_{ij,V/A}(q^2) 
      +  q_{\mu}q_{\nu} 
  \Pi^{[2]}_{ij,V/A}(q^2)
{}  
\end{array} 
\label{correlator}
\end{equation}
with  $m^2 = \sum_{f=u,d,s}  m_f^2$
and  $j^{V/A}_{\mu,ij} = \bar{q}_i\gamma_{\mu}(\gamma_5) q_j$.
Here $q_i$ and $q_j$ are  two (generically different) quarks with
masses $m_i$ and  $m_j$ respectively. In the present paper we work
within QCD with effective three light quarks and do 
not consider corrections due to heavy quarks (c-quark) that can enter
in higher orders of PT through internal loops \cite{chet93}.

An important and convenient property for the analytic analysis  
of the above introduced polarization functions
$\Pi^{[1,2]}_{ij,V/A}(q^2)$
is the absence of the so-called kinematical 
singularities because no
additional factors of momenta appear in the defining
Eq. \re{correlator}. 
For
these polarization functions the  dispersion relations are valid  
that describe  the physical states contributing to the correlator 
\re{correlator} 
\begin{equation} 
\displaystyle
\Pi_{ij,V/A}^{[l]} (q^2) = \frac{1}{12\pi^2}
\int_{s_0}^{\infty}ds
\frac{(-s)^{2 -l} R^{(l)}_{ij,V/A}(s,m_u,m_d,m{},\mu,\alpha_s)}{s-q^2}
\ \ \  \;\;{\rm mod \;sub} 
\label{dispersion.rel}
\end{equation}
where $l=1,2$, and the proper powers of $s$ are introduced 
in the definition of $R(s)$  to
make the spectral densities positive and dimensionless.
In perturbation theory, the threshold is at 
$s_0 = (M_i + M_j)^2$  (where $M_i$ denotes the pole mass of a quark) 
while
the  true thresholds are, say, $4 m^2_\pi$ 
for  $\Pi_{ud}^{[1]}$
and $ m^2_\pi$  for   $\Pi_{ud}^{[2]}$. 
It should be noted that the spectral density
$R^{(2)}$ contains contributions from
spin one as well as  from spin zero intermediate states. 
The spectral density $R^{(0)}$ that is free from contributions
of hadronic states with spin 1
is defined by 
\[
R^{(0)} \equiv  R^{(2)} - R^{(1)} 
{}.
\]
Another useful representation of the tensor $\Pi^{V/A}_{\mu\nu,ij}(q)$ 
  in terms  of scalar functions reads   
\begin{equation}
\Pi^{V/A}_{\mu\nu,ij}(q,m_i,m_j,m{},\mu,\alpha_s)  =
(-g_{\mu\nu}q^2+q_{\mu}q_{\nu}) \Pi^{(1)}_{ij,V/A}(q^2)
                       + q_{\mu}q_{\nu} \Pi^{(0)}_{ij,V/A}(q^2)
{} 
\label{correlator2}
\end{equation}
where the 
correlator is decomposed into the components  
$ \Pi^{(0,1)}_{ij,V/A}(q^2)$
that contains contributions of the states with the angular momentum
$J=0$ and $J=1$ respectively. 

A direct comparison of (\ref{correlator}) and (\ref{correlator2}) leads
us to the following relations 
\beq
\Pi^{(1)}  = -\Pi^{[1]}/q^2, \ \  
\Pi^{(0)}  = \Pi^{[2]} +    \Pi^{[1]}/q^2
{}.
\label{Pi12TOPi01}
\eeq
In  general $\Pi^{[1]}(0)$ may be different from zero 
which implies  a kinematical
singularity (pole) in both  $\Pi^{(1)}$ and  $\Pi^{(0)}$
that describe the spin structure of the correlator
\re{correlator}.

\subsection{Ward identity}
The divergence of the (axial)vector current is known through equations
of motion for the fields in QCD and is given by 
(pseudo)scalar two-quark operators in the case of massive quarks. 
In the massless limit 
both (nonsinglet) axial and vector currents are
conserved. Nevertheless their behaviour is different and, 
for the axial current,  
is governed by the
spontaneous violation of  chiral symmetry and the existence of
the massless excitation accompanying this violation -- the Goldstone
boson,  in our case  the
pion or the kaon.
The (axial)vector and (pseudo)scalar correlators are connected
through a Ward identity 
\begin{equation}
q_\mu q_\nu
\Pi^{{\rm V/A}}_{\mu\nu,ij}(q) = 
(m_i \mp m_j)^2 \Pi^{\rm S/P}_{ij}(q)
+
(m_i \mp m_j)
(
\langle 
\overline{\psi}_{{\rm i}} \psi_{{\rm i}}
\rangle
\mp
\langle 
\overline{\psi}_{{\rm j}} \psi_{{\rm j}}
\rangle 
)
{}\, 
\label{axial-ward}
\end{equation}
where 
\begin{equation}
\displaystyle i \int dx e^{iqx}
\langle
T[\, j^{S/P}_{ij}(x) (j^{S/P}_{ij})^{\dagger} (0)\, ] \rangle
  = \Pi^{S/P}_{i,j}(q) = Q^2 \Pi^{(S/P)}_{i,j}(q)
{}
\label{scalar_correlator}
\end{equation}
and $j^{S/P}_{ij} = \bar{q}_i(i\gamma_5) q_j$, $Q^2=-q^2$.
The dimensionless 
function $\Pi^{(S/P)}_{i,j}(q)$  is related to  the correlator of
(pseudo)scalar currents $\Pi^{S/P}_{i,j}(q)$ by one  power of $Q^2$.

\subsection{Contour integrals and  $\tau$ lepton observables}
The hadronic decay rate of the $\tau$ lepton is obtained by integrating
the absorptive parts of the spectral functions with respect to the
invariant hadronic mass. 
Corresponding to two different tensor decompositions \re{correlator},
\re{correlator2} two different  integral 
representations can be obtained.
The first one displays the structure of hadronic contributions 
classified  according to their spin
\beq      \EQN{4}
\ba{ll}
\dsp
R_{\tau}= &  \dsp R^{(1)}_{\tau} +  R^{(0)}_{\tau} 
\\
=  &
\dsp
\int_0^{M_{\tau}^2}\frac{ds}{M_{\tau}^2}
\left(1-\frac{s}{M_{\tau}^2}\right)^2
\left[\left(1+2\frac{s}{M_{\tau}^2}\right) R^{(1)}(s)
+ R^{(0)}(s)\right]
{}
\ea
\eeq
where
\beq    \EQN{5}
R^{(J)}=|V_{ud}|^2(R^{(J)}_{ud,V}+R^{(J)}_{ud,A})
         + |V_{us}|^2(R^{(J)}_{us,V}+R^{(J)}_{us,A}), \quad J=0,1.
\eeq
The  representation of  the total decay
rate through the absorptive parts of the structure functions 
$\Pi^{[1]}$ and $\Pi^{[2]}$  is simpler from the point of view of
its analytic properties for continuation into the complex plane 
and reads
\beq      \EQN{4b}
R_{\tau}= 
\int_0^{M_{\tau}^2}\frac{ds}{M_{\tau}^2}
\left(1-\frac{s}{M_{\tau}^2}\right)^2
\left[\frac{2 s }{M_\tau^2} R^{(1)}(s)
+ R^{(2)}(s)\right]
{}.
\label{equivalent.repr}
\eeq
Due to the analyticity of  $\Pi^{[1,2]}$ 
in the cut complex
$s$-plane (the absence of singularities away from  the physical
cut, even of kinematical singularities at the origin) 
$R_{\tau}$ can be expressed as the contour integral along a
circle C of the radius
$|s|=M_{\tau}^2$
\beq \EQN{6}
R_{\tau}=6i\pi\int_{|s|=M_{\tau}^2}\frac{ds}{M_{\tau}^2}
\left(1-\frac{s}{M_{\tau}^2}\right)^2
\left[\Pi^{[2]}(s)
-\frac{2}{M_{\tau}^2}\Pi^{[1]}(s)\right]
\label{pi1andpi2}
{}.
\eeq
As the behaviour of  $\Pi^{[1,2]}(s)$ 
along the ``the large circle'' of radius $|s| = M^2_\tau$
is assumed to be reliably evaluated within pQCD, 
representation (\ref{pi1andpi2}) leads to a 
 well-defined pQCD prediction
for $R_\tau$. 
Unfortunately, this  is not true for the  
spin-separated parts. 
Indeed, the  direct use of (\ref{Pi12TOPi01}) leads  to
\beq \label{6b}
\ba{ll}
\dsp
R^{(1)}_{\tau} &= 
\dsp
6i\pi\int_{|s|=M_{\tau}^2}\frac{ds}{M_{\tau}^2}
\left(1-\frac{s}{M_{\tau}^2}\right)^2
\left[ \left(1+2\frac{s}{M_{\tau}^2}\right) \Pi^{(1)}(s)
+\Pi^{[1]}(0)/s
\right]
{},
\\
\dsp
R^{(0)}_{\tau} &= 
\dsp
6i\pi\int_{|s|=M_{\tau}^2}\frac{ds}{M_{\tau}^2}
\left(1-\frac{s}{M_{\tau}^2}\right)^2
\left[\Pi^{(0)}(s)
-\Pi^{[1]}(0)/s
\right]
{}.
\ea
\eeq
where the contribution of the singularity at the origin 
(proportional to $\Pi^{[1]}(0)$) has to be included.
A nonvanishing  value of $\Pi^{[1]}(0)$   
is certainly  a nonperturbative 
constant. Thus, within pQCD we cannot predict 
the decay rates $R^{(1,0)}_{\tau}$ separately.  
In the massless limit $\Pi^{(0)} = 0$  within  perturbation
theory 
and $R^{(0)}$ is saturated
by $\Pi^{[1]}(0)$ corresponding to the massless pion (kaon) pole
for the axial part of the correlator. 

On the other hand, the unknown constant drops out if one
considers moments 
\beq
R^{(1,0)k,l}_{\tau}(s_0) =  
\int_0^{s_0}
\frac{ds}{M_{\tau}^2}
\left(1-\frac{s}{M_{\tau}^2}\right)^{k} 
\left(\frac{s}{M_\tau^2}\right)^l
\frac{d R^{(1,0)}_\tau }{ds}
{},
\label{def:moments}
\eeq
with $k \ge 0, \ \ l \ge 1$. 
(Note that  the moments introduced in \cite{DP} 
are related to  ours as
$
R^{kl}_\tau = R^{(1)k,l}_\tau +  R^{(0)k,l}_\tau $.)

The decay rate 
$R_{\tau}$ may be expressed as the sum of different contributions
corresponding to Cabibbo suppressed or allowed decay modes, vector or
axial vector contributions and the mass dimension of the corrections
\beq \EQN{7}                                                              
R_{\tau} = R_{\tau,V} + R_{\tau,A} + R_{\tau,S}                           
\eeq                                                                      
with                                                                      
\beq \EQN{8}                                                              
\ba{ll} \ds                                                               
R_{V} =                                                                   
& \ds                                                                     
\frac{3}{2}|V_{ud}|^2 \left( 1 + \delta_{0} + \sum_{D=2,4,\dots}        
\delta_{V,ud,{D}} \right),  \\                                          
R_{A} =                                                                   
& \ds                                                                     
\frac{3}{2}|V_{ud}|^2 \left( 1 + \delta_{0} + \sum_{D=2,4,\dots}        
\delta_{A,ud,{D}} \right),  \\                                          
R_{S} =                                                                   
& \ds                                                                     
3|V_{us}|^2 \left( 1 + \delta_{0} + \sum_{D=2,4,\dots}                  
\delta_{us,{D}} \right).                                                
\ea                                                                       
\eeq                                                                      
Here $D$ indicates the mass dimension of the fractional corrections
$\delta_{V/A,ij,{D},}$ and $\delta_{ij,{D}}$ denotes the average of
the vector and the axial vector contributions:
$\delta_{ij,{D}}=(\delta_{V,ij,{D}}+\delta_{A,ij,{D}})/2$.  If 
a decomposition into different spin/parity contributions is made or a
particular pattern of  moments is considered then we will use the
corresponding obvious generalization of (\ref{8}). 
For instance, 
\beq
R^{(1)kl}_{S,V} = a_{kl} \,                                                                   
|V_{us}|^2 \left( 1 + \delta^{kl}_0 + \sum_{D=2,4,\dots}                  
\delta^{(1),kl}_{V,us,D} \right).                                                
\eeq   
and 
\beq
R^{(0)kl}_{S,V} =                                                                   
|V_{us}|^2 \left( \sum_{D=2,4,\dots}                  
\delta^{(0),kl}_{V,us,D} \right).                                                
\eeq
Thus, in our notation we have the relation
\beq
\delta^{kl}_{V,us,{2}}
= a_{kl} \delta^{(1),kl}_{V,us,2} + \delta^{(0),kl}_{V,us,2}
\label{}
{}.
\eeq
\section{Mass Terms in Perturbative QCD}

We would like to stress that inclusion of Cabibbo suppressed modes
into the analysis of observables related to $\tau$ lepton decays gives
not only an additional set of experimental data but open conceptually
new possibilities because the massive piece can be measured in
conjunction with the massless contribution thus providing a strict
normalization and reducing the systematic errors of the experimental
data.  
In
this section we therefore compute mass corrections to the moments of
order ($m^2_s/M^2_\tau$) using first finite order perturbation 
theory and then the
resumed perturbation theory with the evaluation of $\alpha_s$ 
and the quark mass
treated exactly (neglecting  only  unknown higher order  
corrections to the $\beta$-function and the quark mass 
anomalous dimension). 

\subsection{Finite order analysis}
Just like the perturbative predictions for the  massless correlators
also the quark mass corrections for the vector and axial correlators are
identical\footnote{This is strictly true only for the perturbative
contributions.} for the case under consideration with 
$m_i = m_s\neq 0$ and $m_j=m_u=m_d=0$.  
The perturbative prediction for the quadratic mass corrections up to
order $\alpha_s^3$ and for arbitrary quark masses has been presented in
\cite{Wboson} for the transversal piece of the correlator.  
The longitudinal piece is related to
the scalar correlator through the Ward identity (\ref{axial-ward}), viz. 
\begin{equation}
q^4\Pi^{[2]}_{us,V/A} +q^2\Pi^{[1]}_{us,V/A} =
q^4 \Pi^{(0)}_{V/A} =
m_s^2 \Pi^{\rm S/P}+
m_s (\langle \overline{s} s \rangle
\mp
\langle \overline{u}  u  \rangle)
{}.
\label{axial-ward2}
\end{equation}
The vacuum expectation values on the r.h.s.  can be understood
within the framework of perturbation theory and  minimal
subtraction.  
Then formally the last term in \re{axial-ward2}
is of order $m^4$.
Working only within the second order in quark masses,
the vacuum expectation values on the right hand side of
(\ref{axial-ward2}) can be safely discarded
at this point. We return to them in the section for nonperturbative
contributions. 
Thus, the
${\cal{O}}(m_s^2)$ contribution to the longitudinal structure function
$\Pi^{(0)}$ can be taken from Ref.~\cite{gssq} where the massless
scalar correlator has been computed at $\alpha_s^3$.                                                                             
The resulting polarization functions  $\Pi^{[l]}_{V/A}, \ \ l=1,2$                               
is  conveniently represented  in                                        
the form 
\begin{equation}                                                            
(Q^2)^{(l-2)}\Pi^{[l]}_{us,V/A}(q)                                   
=                                                                          
\frac{3}{16\pi^2}\Pi^{[l]}_{V/A,0}(\frac{\mu^2}{Q^2}, \alpha_s)    
+                                                      
\frac{3}{16 \pi^2}
\sum_{D \ge 2} Q^{-D}\Pi^{[l]}_{V/A,D}(\frac{\mu^2}{Q^2},
m_s^2,\alpha_s)      
{}.                                                                         
\label{mass-exp}                                                            
\end{equation}                                                              
Here the first term on the rhs  corresponds  to the massless               
limit while the first term in the sum 
stands for quadratic mass corrections.    
A similar decomposition is assumed for the 
polarization functions $\Pi^{(S/P)}_{us}(q)$. 
The results for both polarization functions with $D=2$ read 
\ice{Attention! these are 
both spectral functions 
with the factor 3/(16Pi^2) factorized }
\begin{eqnarray}
\lefteqn{\Pi^{[1]}_{V,2} = 2 m_s^2\left\{
\rule{0.mm}{6mm}
\right.
 l_{\mu Q}
{+} \frac{\alpha_s}{\pi}
\left[
\frac{25}{4} 
-4  \,\zeta(3)
+\frac{5}{3} l_{\mu Q}
+ l_{\mu Q}^2
\right] }
\nonumber\\
&{+}&\left(\frac{\alpha_s}{\pi}\right)^2
\left[
\frac{18841}{432} 
-\frac{1}{360}  \pi^4
-\frac{3607}{54}  \,\zeta(3)
+\frac{1265}{27}  \,\zeta(5)
\Break
\phantom{+\left(\frac{\alpha_s}{\pi}\right)^2}
+\frac{4591}{144} l_{\mu Q}
-\frac{35}{2}  \,\zeta(3)l_{\mu Q}
+\frac{22}{3} l_{\mu Q}^2
+\frac{17}{12} l_{\mu Q}^3
\right]
\nonumber\\
&{+}&\left(\frac{\alpha_s}{\pi}\right)^3 
\left[
l_{\mu Q}
\left(
\rule{0.mm}{5mm}
\right.
\frac{1967833}{5184} 
-\frac{1}{36}  \pi^4
-\frac{11795}{24}  \,\zeta(3)
+\frac{33475}{108}  \,\zeta(5)
\right. \nonumber \\ &{}&  \left.  
\phantom{\left(\frac{\alpha}{\pi}\right)}
+\frac{4633}{36} l_{\mu Q}
-\frac{475}{8}  \,\zeta(3)l_{\mu Q}
+\frac{79}{4} l_{\mu Q}^2
+\frac{221}{96} l_{\mu Q}^3
\right)
+k^{[1]}_3
\left.
\rule{0.mm}{6mm}
\right]
\left.
\rule{0.mm}{6mm}
\right\}
{},
\nonumber\\
\label{pi1}
\end{eqnarray}

\begin{eqnarray}
\lefteqn{\Pi^{[2]}_{V,2} = - 4 m_s^2\left\{
\rule{0.mm}{6mm}
\right. 
1
{+} \frac{\alpha_s}{\pi}
\left[
\frac{7}{3} 
+2 l_{\mu Q}
\right]}
\nonumber\\
&{+}&\left(\frac{\alpha_s}{\pi}\right)^2
\left[
\frac{13981}{432} 
+\frac{323}{54}  \,\zeta(3)
-\frac{520}{27}  \,\zeta(5)
+\frac{35}{2} l_{\mu Q}
+\frac{17}{4} l_{\mu Q}^2
\right]
\nonumber\\
&{+}&\left(\frac{\alpha_s}{\pi}\right)^3
\left[
l_{\mu Q}
\left(
\frac{14485}{54} 
+\frac{3659}{108}  \,\zeta(3)
-\frac{3380}{27}  \,\zeta(5)
+\frac{1643}{24} l_{\mu Q}
\right.
\Break
\phantom{+\left(\frac{\alpha_s}{\pi}\right)^3l_{\mu Q}}
+\frac{221}{24} l_{\mu Q}^2
\right)
\left.
\left.
+k^{[2]}_3
\rule{0.mm}{6mm}
\right]
\right\}
{}.
\nonumber\\
\label{pi2}
\end{eqnarray}

Here  $l_{\mu Q} = \, {\mathrm{ln}}\frac{\mu^2}{Q^s}$,
the  mass $m_s$ as well as QCD coupling constant $\alpha_s$ are
understood 
to be taken at a generic value of the t' Hooft~mass $\mu$. All              
correlators are  renormalized within                                        
${\overline{{\mbox{MS}}}}$-scheme.
Note that terms of order $\alpha_s^3$ are known only with
``logarithmic'' accuracy, that is the constant  parts $k_3^{[1]}$
and  $k_3^{[2]}$ in the large $Q$
behavior of the corresponding correlators are not
available\footnote{In fact the very calculation of the constant parts
is well beyond the present calculational techniques.}.

This means that
we do not know the $O(m_s^2\alpha_s^3)$ 
contributions to $R_\tau$; however
these constants (as well as the also unknown ``low-energy'' constant
$\Pi^{[1]}(0)$) do  not appear in the moments $R^{(1,0)k,l}_{\tau}$
with $k \ge 0, \ \ l \ge 1$.

Let us discuss some concrete results (neglecting for the moment any
nonperturbative condensate contributions and quartic mass corrections 
as well). 
Since the perturbative
result, in particular the $m^2$ terms,  do not differentiate between vector
and axial vector channels we will consider in the following 
their sum. 
The mass corrections to the moments of spin $0$ and spin
$1$  final state distributions can now be cast into the following form
($m_s = m_s(M_\tau)$, $\alpha_s = \alpha_s(M_\tau)$)
\beq
\delta^{kl}_{us,2} = 
b_0\frac{m_s^2}{M_\tau^2}
\{
1 
+ b_1 \frac{\alpha_s}{\pi} 
+ b_2 \left(\frac{\alpha_s}{\pi}\right)^2 
+ b_3 \left(\frac{\alpha_s}{\pi}\right)^3 
\}
{}.
\label{form1}
\eeq
In view of the large size  of the  coefficients 
$b_i$
we attempt to reduce the higher order corrections 
by adopting the value $1$ GeV  for the renormalization  scale
of the running mass and define an alternative  set of coefficients 
$\hat{b}_i$ through
\beq
\delta^{kl}_{us} = 
\hat{b}_0\frac{m_s^2(1 \ \ \mbox{GeV}) }{M_\tau^2}
\{
1 
+ \hat{b}_1 \frac{\alpha_s}{\pi} 
+ \hat{b}_2 \left(\frac{\alpha_s}{\pi}\right)^2 
+ \hat{b}_3 \left(\frac{\alpha_s}{\pi}\right)^3 
\}
{}.
\label{form2}
\eeq
The results are listed in  Table 1. 
The same expansions can also be obtained for the correction to the spin
zero and spin one parts separately. In view of the additional 
nonperturbative contribution $\sim \Pi^{[1]}(0)$ which appears
in the lowest order moment ($ l=0 $)  due to the spin separation 
only the results for $ l \ge 1$ are given in Tables 2-3 for these 
latter cases. 
The additional 
nonperturbative contribution $\sim \Pi^{[1]}(0)$ in the axial vector
current can be estimated 
in the chiral limit and happens to be connected
to the contribution of the pseudoscalar meson (pion or kaon)
to the correlator. The distinguished
role of the pion is due to its nature as a Goldstone particle
related to spontaneous violation of the chiral symmetry of QCD.
When the explicit violation of the chiral symmetry given by 
nonzero 
values of quark masses is small these masses provide small
corrections to the massless limit (pion dominance) that can be
accounted for on a regular basis within Chiral Perturbation Theory \cite{GL}.
As for corresponding quantity in the vector channel there is no solid
physical arguments for estimating its value with finite masses of
quarks though it vanishes in the massless limit due to vector current
conservation and the pattern of spontaneous symmetry 
breaking.  

\subsection{Resummation of effects of running}

The rapid  change of coefficients of perturbation theory expansions 
for different moments is 
caused by the running of the coupling constant and the mass along the
contour of integration. The resummation of these effects can be
performed in all orders of $\alpha_s$.
The technique for the massless case is described in the literature 
\cite{tau:resum,DP,resum}
so here we concentrate on the massive case. It introduces no much new
technically but extends the freedom of choosing the resummation
procedure due to additional parameter -- the running mass of a quark.

Let us, nevertheless,  recall the central idea  for the  massless case. 
The main object to start with is the Adler's $D$-function 
for   the transverse part
of the correlator.
The orders of perturbation theory in 
the $\overline{\rm MS}$ scheme are formally counted 
after the renormalization group improvement of the series.
The information 
from the perturbative treatment  
is contained in the string of the coefficients of powers
of the running coupling.
The polarization function that appears in the integral over the large
circle is then restored from the $D$-function
(including renormalization group 
improvement)
by solving the corresponding
evolution 
equation exactly with the $\beta$ function taken to a given fixed
order in $\alpha_s$.
Let us demonstrate the procedure in a simple example.
Consider a $D$-function with its 
leading term normalized to unity
(here and below we are using 
$a\equiv \alpha_s/\pi$)
\begin{equation}
  \label{dfunc}
D(Q^2)=1+a(Q^2)+k_1a(Q^2)^2+k_2a(Q^2)^3+k_3a(Q^2)^4+\ldots  
{}.
\end{equation}
It is  connected to   $\Pi(Q^2)$ 
   by
\begin{equation}
  \label{dequa}
D(Q^2)=-Q^2\frac{d}{d Q^2}\Pi(Q^2) 
{}.
\end{equation}
In  leading order for the $\beta$ function
\begin{equation}
  \label{betafunc}
  \beta(a)=-\beta_0 a^2, \quad  Q^2\frac{d}{d Q^2}a(Q^2)=  \beta(a(Q^2))
\end{equation}
the   polarization function 
can be completely restored in the closed  form.
Direct integration of Eq. \re{dequa} gives
the improved polarization function
\begin{eqnarray}
  \label{imppol}
\Pi(Q^2)&=&\ln (\mu^2/Q^2) +\Pi(\mu^2)\nonumber \\
&+& \frac{1}{\beta_0}\left(   
-\ln\frac{1}{a(Q^2)}+k_1a(Q^2)+\frac{k_2}{2}a(Q^2)^2+
\frac{k_3}{3}a(Q^2)^3+\ldots  \right)
\end{eqnarray}
that ought to be integrated along the contour in 
the complex plane with 
\begin{equation}
  \label{1ordersol}
a(Q^2)={a(\mu^2)\over 1+\beta_0 a(\mu^2) \ln(Q^2/\mu^2)}  
\end{equation}
which can be continued into the complex $Q^2$ plane. The normalization
point $\mu$ is conveniently chosen to be equal to the $\tau$ lepton
mass
$\mu=M_\tau$. 
The part independent of $Q^2$ (the integration constant $\Pi(\mu^2)$)
does not contribute to the integral. This is also obvious from the spectral
integral itself because  $\Pi(\mu^2)$ being a constant has no
discontinuity across the physical cut in  the $Q^2$ complex plane.
The first term in \re{imppol} is simply the partonic contribution 
$\ln (\mu^2/Q^2)$.

The generalization to higher orders of $\beta$ function is
straightforward. In second order the integration of Eq. \re{dequa} can
be still performed explicitly while for third and fourth order of the
$\beta$ function numerical integration is more convenient.

To make things clearer let us show the 
explicit
example in the second order for the $\beta$ function
\begin{equation}
  \label{beta2}
  \beta(a)=-\beta_0 a^2-\beta_1 a^3  .
\end{equation}
(In higher orders the $\beta$-function is  scheme dependent.
However, there is always
a scheme where this is the complete expression for the $\beta$
function
 -- the t' Hooft scheme. Then the convention could be to resum always in
this scheme that is justified by technical simplicity. 
Nevertheless, we 
 however
stick to the $\msbar$- scheme in the present paper.)
A contribution to the $D$-function of the form 
$\Delta D(Q^2)=a(Q^2)^2$  leads after integration of
Eq. \re{dequa} to the corresponding 
contribution to the polarization function 
$\Delta \Pi(Q^2)$ of the
form 
\begin{equation}
  \label{pol2}
\Delta \Pi(Q^2)=\frac{1}{\beta_1}\ln\left(a(Q^2)+\frac{\beta_0}{\beta_1}
\right)+\Delta \Pi(\mu^2)
\end{equation}
as  can be easily checked by direct differentiation.
The improved  polarization function $\Delta \Pi(Q^2)$ 
\re{pol2} can then be used for integration along the contour.
The remark about the constant terms applies  again.

The appearance of mass introduces another freedom in the choice of
the basic quantities that accumulate the perturbative information.
The actual procedure is described below. 
As basic quantities we choose
$ \Pi^{[1]}_{us}(q^2)  $ and 
$ \Pi^{[2]}_{us}(q^2)$.
The  renormalization of the pieces proportional to $m_s^2$
is  different for 
$ \Pi^{[1]}  $ and 
$ \Pi^{[2]}$.
The second one $ \Pi^{[2]}_{us,2}(Q^2)$, 
is scale-invariant  and the renormalization group
improvement can be performed  directly
\begin{equation}
  \label{piq}
Q^2 \Pi^{[2]}_{us,2}(Q^2) =  k^{[2]}_0 m_s^2(Q^2)(1 +k^{[2]}_1 a(Q^2)+\ldots) .
\end{equation}
In contrast $\Pi^{[1]}$ is
not renormalized multiplicatively and the corresponding renormalization group
equation is not uniform i.e it has a free term. 
The problem is solved by 
introducing the corresponding $D$-function $D^{[1]}(Q^2)$
by one differentiation with respect to $Q^2$.
The result is 
\begin{equation}
  \label{gpart}
D^{[1]}_{us,2}(Q^2)= -\frac{Q^2}{2}\frac{d}{d Q^2} \Pi^{[1]}_{us,2}(Q^2)
=  m_s^2(Q^2)(1 + d^{[1]}_1 a(Q^2)+\ldots).      
\end{equation}
Then we proceed as explained before in construction of  the polarization 
operator.
The running of the mass as taken into account 
through the renormalization group equation
\begin{equation}
  \label{mrg}
 Q^2\frac{d}{d Q^2}m_s(Q^2)=  \gamma_m(a(Q^2))m_s(Q^2) , \quad  
\gamma_m(a)=-\gamma_0 a-\gamma_1 a^2-\ldots
\end{equation}
with the solution 
\begin{equation}
  \label{rgmasssol}
m_s(Q^2)=m_s(\mu^2)\exp\int_{\alpha_s(\mu^2)}^{\alpha_s(Q^2)}
{\gamma(x)dx\over \beta(x)}  
{}.
\end{equation}
Subsequently, the integration along the contour can be performed 
directly.

The explicit formula  in the leading order of 
the $\beta$ function for $\Pi^{[2]}(Q^2)$
is easily found.
Having solved the Eq. \re{mrg} for the leading order $\beta$ and
$\gamma$ functions (or using the explicit formula \re{rgmasssol})
we have 
\begin{equation}
  \label{piqimp}
Q^2 \Pi^{[2]}_{us,2}(Q^2) 
= k^{[2]}_0 m_s^2(\mu^2)\left(a(Q^2)\over
  a(\mu^2)\right)^\frac{2\gamma_0}{\beta_0}
(1 + k^{[2]}_1 a(Q^2)+\ldots) 
\end{equation}
where $a(Q^2)$ is the solution to the renormalization group 
Eq.~\re{1ordersol}.

For the second amplitude the result  is  slightly more complicated,
namely
\[
\Pi^{[1]}_{us,2}(Q^2) 
\]  
\begin{equation}
  \label{pigimp}
=-{m_s^2(\mu^2)\over \beta_0}
\left(a(Q^2)
\over
  a(\mu^2)\right)^\frac{2\gamma_0}{\beta_0}
\left({1\over{\left( \frac{2\gamma_0}{\beta_0}-1\right)}}\frac{1}{a(Q^2)}
+{d^{[1]}_1\over \frac{2\gamma_0}{\beta_0}}
+{d^{[1]}_2\over \frac{2\gamma_0}{\beta_0}+1}a(Q^2)
+\ldots\right).
\end{equation}

These formulae should be substituted in \re{pi1andpi2},\re{6b} and
integrated along the contour.  The generalization to higher orders of
the $\beta$ and $\gamma$ functions is straightforward.  Again up to
the second order expansion of these functions in $\alpha_s$ explicit
analytical integration can be performed in terms of elementary
functions for some amplitudes.  (Third order also allows some
integrations in terms of elementary functions but formulas become too
awkward and the numerical treatment is preferable.) The coefficients
$d^{[1]}_i$ can be inferred from the explicit expression for the
polarization function \re{pi1}.  They accumulate the whole
perturbative information because the $\beta$-function and the quark
anomalous dimension necessary for the restoration of the full
expressions \re{pi1} and
\re{pi2} are known.

We perform the analysis along these lines for the same moments as in
the finite order of perturbation theory with the available
coefficients for $\Pi^{[2]}$ and for the $D$ functions in the case of
$\Pi^{[1]}$ and with $\beta$ and $\gamma$ functions from first to
fourth order \cite{beta4,gamma4,gamma4full}.

For the presentation it is convenient to order the results according to
the corresponding contributions from $D/\Pi$ functions and to
attach a power of $\alpha_s(M_\tau^2)$ to every consequent term as a
mark.

For instance the typical answer analogous to Eqs. \re{form1},\re{form2}
can then  be cast into the form
(here and below $m_s = m_s(M_\tau)$) 
\beq
 \tilde{\delta}^{kl}_{us}  = 
\tilde b_0(\alpha_s)\frac{m_s^2}{M_\tau^2}
\{
1 
+ \tilde b_1(\alpha_s) \frac{\alpha_s}{\pi} 
+ \tilde b_2(\alpha_s) \left(\frac{\alpha_s}{\pi}\right)^2 
+ \tilde b_3(\alpha_s)\left(\frac{\alpha_s}{\pi}\right)^3 
\}
\label{form1imp}
\eeq
where the coefficients 
$\tilde b_i(\alpha_s)$
are obtained with resummation and become functions of $\alpha_s$.
These functions are not given in an explicit form but are fixed
numbers obtained from the actual analysis because the contour
integration was not done in general analytic form but only
numerically. The resummation procedure is expected to exhibit an
improved behaviour of the series compared to the finite order analysis
and to provide a definite coefficient in front of the free parameter $m_s^2$.

The results are presented in Tables 3-6.
By comparing the results of these two analyses one can judge the 
degree of improvement achieved by resummation in 
${\overline{\rm MS}}$ scheme.
\section{Non-perturbative corrections for $J=0$ moments}
Besides the perturbative radiative corrections to $R_{\tau}$ also    
nonperturbative QCD effects influence the hadronic $\tau$ decay
rate and its differential distributions analyzed with the moments. 
The short distance operator product expansion (OPE) for the 
spectral functions  
\beq \EQN{t1}                                                    
\Pi^{(J)}(s=-Q^2) = \sum_{D=0,2,4,\cdots}\frac{1}{(Q^2)^{D/2}}       
               \sum_{{\rm dim}{\cal O} = D} C^{(J)}(Q^2,\mu)      
               \langle{\cal O}(\mu)\rangle                        
\eeq                                                              
may be used to take into account both perturbative and 
nonperturbative contributions. 
Here we collect the known non-perturbative corrections
to the structure function $\Pi^{(0)}_{V/A}$.

\subsection{Dimension $D=4$ and  $D=6$ Corrections}
Due to  Ward identity \re{axial-ward2}
the $D=4$ part of the  polarization  function $\Pi^{(0)}_{V/A}$ can be cast
into the following convenient form
\beq
\Pi^{(0)}_{V/A,4}(Q^2)  = 
\frac{m_s^2 \Pi_{(S/P),2}(Q^2)}{Q^4}
+
\frac{
m_s 
(
\langle
\bar s s
\rangle
\mp
\langle
\overline{u}  u 
\rangle
)}{Q^4}
\frac{16\pi^2}{3}
{}
\label{axial-ward3}
\eeq
where  $\Pi_{(S/P),2}$ is 
\cite{ms:as^3-analysis}
\begin{eqnarray}
\lefteqn{\Pi_{S/P,2} =  {m_s^2}\left\{
\rule{0.mm}{4mm}
\right.
 \nonumber
-4 
-4 l_{\mu Q} 
}
\nonumber\\
&{+}& \frac{\alpha_s}{\pi}
\left[
-\frac{100}{3} 
+16  \,\zeta(3)
-\frac{64}{3} l_{\mu Q}
-8 l_{\mu Q}^2
\right]
\nonumber\\
&{+}&\left(\frac{\alpha_s}{\pi}\right)^2
\left[
-\frac{33109}{108} 
+\frac{2}{135}  \pi^4
+\frac{1978}{9}  \,\zeta(3)
-\frac{620}{9}  \,\zeta(5)
\Break
\phantom{+\left(\frac{\alpha_s}{\pi}\right)^2}
-\frac{5065}{18} l_{\mu Q}
+\frac{308}{3}  \,\zeta(3)l_{\mu Q}
-97 l_{\mu Q}^2
-\frac{50}{3} l_{\mu Q}^3
\right]
\left.
\rule{0.mm}{6mm}
\right\},
\label{D=4}
\\
&=&
-4\left(
1. + 3.5251\frac{\alpha_s}{\pi}  + 28.0923 \left(\frac{\alpha_s}{\pi}\right)^2
\right)
{}.
\label{D=4:N}
\end{eqnarray}
In the last line  we have set 
$\mu^2 = Q^2$.
\ice{
                             100    64 LMQ        2
Out[125]= -4 - 4 LMQ + as (-(---) - ------ - 8 LMQ  + 16 z3) + 
                              3       3
    }
Note that there are no  
corrections  to the coefficient function of the
quark condensate term in Eq.~(\ref{axial-ward3}) 
because it comes from equal-time
commutation relation for the quark fields
that are just initial conditions for the
field theory. 

It is clear from \re{axial-ward2} that these power corrections
are essentially given by the corresponding $D=4$ terms of the
OPE for the  structure function $\Pi^{(S)}_{V/A}$ which are  well-known
(see, e.g. Refs. \cite{ms:Jamin95,ms:as^3-analysis} and references therein). 
As a result we 
have
\bea
\frac{3}{16\pi^2}\Pi^{(0)}_{V/A,6}  = 
m_s^2 \left[\rule{0mm}{6mm}
\right.
&+ & \frac{1}{8}
\left(
 1 + \api\frac{11}{2} + 2 \api l_{\mu Q}
\right) \api \langle G G\rangle
\\
\nonumber
& \pm & 
\left( 1 + \frac{14}{3}\api +2 \api l_{\mu Q}
\right) 
m_s \langle\bar{u} u\rangle 
\\
\nonumber
& + &
\frac{1}{2}\left( 1 + \frac{11}{3}\api + 2 \api   l_{\mu Q}
\right) 
m_s \langle\bar{s} s\rangle 
\\
\nonumber
& + &
\frac{3}{16\pi^2}
m_s^4
\left( 1 
- 2 l_{\mu Q}
+ 
\api \left[
8 \zeta(3) - 6 
- 4 l_{\mu Q}
- 8 l_{\mu Q}^2
\right] 
\right) 
\left.
\rule{0mm}{6mm}
\right]
{}.
\label{Pi0m6}
\eea
The corresponding contributions  to the $(k,l) = (0,1)$ moment
read (to save space we have partially converted 
the coefficient in front of $m_s^4$ term to numbers 
and put $\mu= M_\tau$ so that
below $\alpha_s = \alpha_s(M_\tau)$ and $m_s = m_s(M_\tau)$).
\beq
\delta^{(0),01}_{us,4}
= \frac{45}{2} \frac{m_s^4}{M_\tau^4}
\left[1. + 4.77815 \api + 31.0478 \left( \api \right)^2 \right]
-12 \pi^2 
\frac{\dsp
\left(
\langle m_s\bar{s}s\rangle \mp \langle m_s\bar{u}u\rangle
\right)
     }{m_s^4}     
\label{D=4:moment01}
{}
\eeq
and
\bea
\delta^{(0),01}_{us,6}  = 
\frac{m_s^2}{M_\tau^6} \left[\rule{0mm}{6mm}
\right.
&-3& \pi^2
\left(
 1 + \api\frac{11}{2} 
\right)\langle \frac{\as}{\pi}GG\rangle
\nonumber
\\
\nonumber
& \mp & 24 
 \pi^2
\left( 1 + \frac{14}{3}\api 
\right) 
 \langle m_s\bar{u}u\rangle
\\
\nonumber
&-&
12 \pi^2
\left( 1 + \frac{11}{3}\api 
\right) 
\langle m_s\bar{s}s\rangle
\\
&-&
\frac{9}{2} m_s^4 
\left( 1 
+
\api \left[
8 \zeta(3) + \frac{8}{3}\pi^2 - 22 
\right] 
\right) 
\left.
\rule{0mm}{6mm}
\right]
{}.
\label{D=6:moment01}
\eea
Note, please, that in Eqs.~(\ref{D=4:moment01},\ref{D=6:moment01}) 
we have used  the very quark and gluon
condensates normalized at the ``natural'' scale $\hat{\mu}=M_{\tau}$
for our RG invariant condensates (for more details, see 
\cite{Chetyrkin93,Chetyrkin93-update,ms:as^2-analysis}):
\beq \EQN{t3b}
\ba{ll} \dsp
\langle \frac{\as}{\pi}GG\rangle = &  \dsp
 \langle\frac{\as(M_{\tau})}{\pi} GG({\mu}=M_{\tau})\rangle, 
\\ \dsp
\langle m_i\bar{\Psi}_j\Psi_j\rangle = &  \dsp
\langle m_i(M_{\tau})\bar{\Psi}_j\Psi_j({\mu}=M_{\tau})\rangle
.\ea\eeq

\section{Discussion of numerical results}
Let us first consider the mass corrections to the total rate
within fixed order perturbation theory.  The
coefficients $b_i$ are already fairly large for all the moments.  Even
worse, however, is the fact that the series grows dramatically for all
the moments, rendering a mass determination from the total rate or the
corresponding momenta potentially  unreliable. The transition from
$m_s(M_\tau)$ to $m_s(1~{\rm GeV})$ reduces the coefficients only
marginally.  From Table 2 it is evident that the large corrections
originate mainly from the spin zero contribution to the moments,
leaving significantly smaller QCD corrections to the spin one
part. Hence, it is in principle possible to determine
$m_s$ from the spin one part alone. Alternatively one might try to 
combine spin zero and spin one moments with different relative
coefficients to artificially decrease the QCD corrections further.
However, in the absence of any physically motivated reasonings 
we refrain from such an approach. 
The main problem of the strange quark determination and even the
total perturbative analysis of the Cabibbo suppressed $\tau$ lepton
decays is the interpretation of the perturbation theory series for
$m_s^2$ corrections. In passing we note that also perturbative
corrections to the power suppressed terms proportional to
$m_s^4$ and $m_s^6$  
(see Eqs.~(\ref{D=4:moment01},\ref{D=6:moment01}))
might  provide us with another example
of bad behaviour of higher order terms. However, since
the ${\cal O }(\alpha_s^2$ corrections to the spin one (and,
as a consequence, to the total $R_\tau$) are available we concentrate
below  at discussing the $m_s^2$ terms only.

The problem is fairly obvious from a consideration of the two lowest
order moments. Let us use $\alpha_s(M_\tau) = 0.334$ 
\cite{hocker} 
in computing the
size of the perturbative contributions. 
The order of magnitude of the unknown coefficient $k_3^{[2]}$
in  Eq.~(\ref{pi2})
is estimated on the basis of a geometric series to amount to
$k_3^{[2]} =  0 \pm 19.6^2/2.33 
= 0 \pm  (k_2^{[2]})^2/k_1^{[2]} = 0 \pm 164.4$.
For  fixed order  we thus find series is  thus
given by (below $a= \alpha_s(M_\tau)/\pi$ and $m_s = m_s(M_\tau)$)
\bea
{\delta}^{00}_{us,2} &=& 
-8\frac{m_s^2}{M_\tau^2}(
1. + 5.33  a + 46.0  a^2  + 284  a^3  + 0.75 a^3  k_3^{[2]} 
  ) 
 \nonumber \\ 
&{=}&
-8\frac{m_s^2}{M_\tau^2}(1. + 0.567  + 0.520   + 0.341 \pm  0.148 )
 \nonumber \\ 
&{=}&
-8\frac{m_s^2}{M_\tau^2}
( 2.4 \pm 0.5)
\label{num:disc:1.3}
{},
\eea
where we have assumed the (maximal!) value of the ${\cal{O}}(\alpha_s^3)$
term as an estimate of the theoretical  uncertainty
(this convention will  be used also below).
For the  ``contour improved'' series one obtains 
\bea
\tilde{\delta}^{00}_{us,2} &=& 
-8\frac{m_s^2}{M_\tau^2}(
1.44 + 3.65  a + 30.9 a^2  + 72.2  a^3  + 1.18  a^3 k_3^{[2]} 
  ) 
 \nonumber \\ 
&{=}&
-8\frac{m_s^2}{M_\tau^2}(1.44 + 0.389 + 0.349   + 0.0867   \pm 0.234)
 \nonumber \\ 
&{=}&
-8\frac{m_s^2}{M_\tau^2}
( 2.26 \pm  0.32)
\label{num:disc:2.3}
{}.
\eea
Comparing the ``improved'' with the finite order analysis one
observes that higher orders  give numerically smaller contributions 
although  the apparent  convergence is rather marginal as well.
The ratio between two correction terms 
$\tilde{\delta}^{00}_{us,2}/{\delta}^{00}_{us,2} \approx 0.95$
shows a relatively stable behaviour.
The  results within resummation technique are stable within the allowed
range of $\alpha_s=0.312-0.356$ as  checked by direct computation.
\ice{N1:}
Although no firm prediction can be made as a consequence of the large 
corrections, it  is at least encouraging to observe that the sign of 
the mass correction  remains preserved and the consecutive terms 
exhibit a marginal decrease. 
Eq.~(\ref{num:disc:2.3}), with the  uncertainty increased  by perhaps
a factor 2 could be considered as a reasonable estimate of the strange 
mass corrections.

In fixed order approximation the moments  with $l \ge 1$ are independent 
of the constant $k_3^{[2]}$;   a residual dependence remains,
 however,
in the  ``contour improved'' treatment
\bea
{\delta}^{01}_{us,2} &=& 
-\frac{5}{3} \frac{m_s^2}{M_\tau^2}( 
1 - 4.17  a - 113.  a^2  - 1820. a^3
                       ) 
 \nonumber \\ 
&{=}&
-\frac{5}{3}  \frac{m_s^2}{M_\tau^2}
(1 - 0.443  - 1.27   - 2.19 )
 \nonumber \\ 
&{=}&
-\frac{5}{3}  \frac{m_s^2}{M_\tau^2}
                     (             
-2.9 \pm 2.2
                     )
\label{num:disc:3.3}
{},
\eea
and
\bea
\tilde{\delta}^{01}_{us,2} &=& 
-\frac{5}{3} \frac{m_s^2}{M_\tau^2}(
 -2.26 - 14.7  a - 204.  a^2  - 171.  a^3  - 13.3  a^3  k_3^{[2]}
                         ) 
 \nonumber \\ 
&{=}&
-\frac{5}{3} \frac{m_s^2}{M_\tau^2}( 
-2.26 - 1.56 - 2.3   - 0.206   \mp 2.62 
  )
 \nonumber \\ 
&{=}&
-\frac{5}{3} \frac{m_s^2}{M_\tau^2}
( -6.3 \pm 2.8 )
\label{num:disc:4.3}
{}.
\eea
The $(0,1)$ moments 
\ice{ ${\delta}^{01}_{us,2}$  }
thus  
exhibit  a rapid growth of the coefficients and, at the  same time,
with  $\tilde{\delta}^{01}_{us,2}/{\delta}^{01}_{us,2} =2.2$
a strong dependence on the improvement procedure.
This comparison shows that
there is no consistent prediction in $\msbar$ scheme for 
this observable --
the first moment of the differential rate.

Now we turn to  to the contributions of the spin one and spin zero
separately.
As stated before, the lowest moments $(l=0)$ 
of the spin-dependent functions depend on the nonperturbative quantity
$\Pi^{[1]}(0)$. 
For the spin one part and for 
$(k,l) = (0,1)$  we find
\bea
{\delta}^{(1)01}_{us,2} &=& 
-5 \frac{m_s^2}{M_\tau^2}( 
1. + 4.83  a + 35.7  a^2  + 276. a^3
                       ) 
 \nonumber \\ 
&{=}&
-5  \frac{m_s^2}{M_\tau^2}
(1. + 0.514  + 0.404   + 0.331 )
 \nonumber \\ 
&{=}&
-5 \frac{m_s^2}{M_\tau^2}
(2.25 \pm 0.33)
{}
\label{num:disc:5.3}
\eea
and
\bea
\tilde{\delta}^{(1)01}_{us,2} &=& 
-5\frac{m_s^2}{M_\tau^2}(
1.37 + 2.55  a + 16.1  a^2  + 135 a^3 ) 
 \nonumber \\ 
&{=}&
-5 \frac{m_s^2}{M_\tau^2}(  1.37 + 0.271  + 0.182   + 0.163  )
 \nonumber \\ 
&{=}&
-5  \frac{m_s^2}{M_\tau^2}
(  2.0  \pm  0.2)
{}.
\label{num:disc:.6.3}
\eea
Note that spin 1 contribution is determined by the component 
$\Pi^{[1]}$ alone and is known up to third order.
Clearly, this series is decreasing in a reasonable way
(comparable to the behaviour of  $\tilde{\delta}^{00}_{us,2}$)
 and, at the same
time, only moderately  dependent on the improvement 
prescription with
$\tilde{\delta}^{(1)01}_{us,2}/{\delta}^{(1)01}_{us,2} = 0.89$.
On the basis of Eq.~(\ref{num:disc:.6.3})  
this moment might well serve for a reliable $m_s$ determination, with 
a sufficiently  careful interpretation of the theoretical
uncertainty.  

The corresponding spin zero part is, per se, proportional to 
$m_s^2$ (not counting the ``condensate'' contributions discussed in 
Section 4) and thus could be considered as ideal for a 
measurement  of $m_s$. However, the behaviour of the perturbative
series 
\bea
{\delta}^{(0)01}_{us,2} &=& 
\frac{3}{2} \frac{m_s^2}{M_\tau^2}( 
1. + 9.33  a + 110 a^2  + 1323 a^3) 
\nonumber \\ 
&{=}&
\frac{m_s^2}{M_\tau^2}
(1. + 0.992  + 1.24   + 1.59 )
 \nonumber \\ 
&{=}&
\frac{3}{2} \frac{m_s^2}{M_\tau^2}
(4.8 \pm 1.6 )
\label{num:disc:.7.3}
{}
\eea
and
\bea
\tilde{\delta}^{(0)01}_{us,2} &=& 
\frac{3}{2}\frac{m_s^2}{M_\tau^2}(
3.19 + 11.2  a + 126.  a^2  + 289.  a^3  + 6.63  a^3  k_3^{[2]}
                         ) 
 \nonumber \\ 
&{=}&
 \frac{3}{2}\frac{m_s^2}{M_\tau^2}( 
3.19 + 1.19  + 1.42  + 0.347    \pm 1.31
                                  )
 \nonumber \\ 
&{=}&
\frac{3}{2}  \frac{m_s^2}{M_\tau^2}
( 6.14 \pm 1.6 )
\label{num:disc:.8.3}
{}
\eea
shows  a rapid growth of the coefficients. The series is not expected
to provide an accurate prediction for the mass effects. (The same 
rapid growth of the coefficients in the perturbative series
is present for  the scalar correlator related to 
$\Gamma(H \to b \bar{b})$ \cite{gssq}.) 
Nevertheless, an interesting estimate could also be deduced  from  the spin
zero contributions, in particular after resummation. For the spin zero, 
spin one and the total  rate the following relation holds
\[
{\delta}^{01}_{us,2} = {\delta}^{(1)01}_{us,2} 
+\frac{20}{9} {\delta}^{(0)01}_{us,2}.
\]

Besides the observables themselves one has also look at convergence of
the  $\beta$ and $\gamma$ functions that determine the running along the
contour.
In the present  case 
\[
\beta(a)=-2.25 a^2(1+1.78 a+ 4.47 a^2+ 21.0 a^3)
\]
or at $a=0.1$
\[
\beta(0.1)=-0.0225
(1+0.18 + 0.045+ 0.021)
\]
which is quite  good.
For the $\gamma$ function, however, 
\[
\gamma(a)=- a(1+3.79 a+ 12.4 a^2+ 44.3 a^3)
\]
and the convergence is marginally acceptable.

The resumed series behave in general better than those of finite
order.
However,  for the mass corrections  they still do not 
satisfy the heuristic criteria 
of convergence.
In  practice the resummation  maintains  the convergence 
pattern of the corresponding $D$-function.  For
the  mass corrections the  $D$-functions themselves exhibit 
rapidly growing coefficients of the perturbative series,
whence the resummation does not lead to a significant  
improvement.  Numerically the
convergence for $D$-functions in $\overline{\rm MS}$ scheme is 
marginal. For the present case  the $D$-function for
$\Pi^{[1]}(Q^2)$ is given by
\[
 D^{[1]}(Q^2)= m_s^2(Q^2)(1+\frac{5}{3}a+a^2
\left(\frac{4591}{144} - \frac{35}{2}\zeta(3)\right)
\]
\[
+a^3
\left(\frac{1967833}{5184} - \frac{\pi^4}{36} 
- \frac{11795}{24}\zeta(3) + \frac{33475}{108}\zeta(5)\right)
\]
\[
= m_s^2(Q^2)(1+1.67 a + 10.84 a^2+107.53 a^3)
\]
whereas  a  much better pattern
of convergence is observed for massless part
\[
  D_0(Q^2)=1+a+a^2
\left(\frac{299}{24} - 9\zeta(3)\right)
+a^3\left(\frac{58057}{288} - 
\frac{779}{4}\zeta(3) + \frac{75}{2}\zeta(5)\right)
\]
\[
=1+a+1.64 a^2 + 6.37 a^3 \,.
\]
This  is the reason why the
precise determination of the strong coupling constant from the $\tau$
lepton lifetime is possible.

\section{Summary}
Tau decays in the $\Delta S =1$ channel are sensitive to the mass of the 
strange quark.  Moments of the four spectral functions corresponding 
to spin zero and spin one contributions, induced by 
vector and axial vector currents can be evaluated in perturbative QCD.
With this motivation we have  evaluated the QCD corrections
up to order $\alpha_s^2$ for the  rate and up to order 
$\alpha_s^3$ for the higher moments  of the spectral function, separated
according  to the spin zero and spin one contributions. 
Resummation of the RG enhanced terms  has been applied to the 
various monets.

At first glance, 
the  spin zero part of the rate and of the moments appears
to be particularly promising for an  $m_s$  determination, since
this piece is multiplied by $m_s^2$ in the parton model. However,
as a consequence of the spin zero separation  only moments with $l >0$ 
can be calculated in pQCD. In addition, it turns out that the  
perturbative corrections to the  spin zero part are remarkably
large, invalidating the straightforward extraction of $m_s$ from such 
an analysis. 

Also the total rate, i.e. the combination of spin zero and spin one
contributions, exhibits a small mass dependent term which can, in
principle, also be used to determine $m_s$. However, again, one
observes remarkably large QCD corrections which can be traced---at
least for the higher moments---to the spin zero part. This is clearly
visible by considering the mass terms of the separated spin one piece,
which exhibit reasonably smaller coefficients of the first, second and
the third order corrections.

Resummation leads to a modest improvement of the apparent convergence
of the series, such that the lowest moments can be predicted with
acceptable accuracy.  Assuming sufficiently precise data, a combined
analysis of the spin zero and one moments might thus lead to a
reliable determination of the strange quark mass.

We do not discuss $\mu$ (renormalization scale) dependence in $\msbar$
scheme as a special topic fixing it always at $\mu=M_\tau$.  Our
estimate of the error for ``contour improved'' analysis ($2.26\pm
0.6$) is rather conservative and covers all reasonable change of
$\mu$. The more general topic of investigating the entire scheme
dependence of this quantity is outside the scope of the paper.

To conclude  we note that the large value for the correction factor
$\tilde{\delta}^{00}_{us,2}/(-8)$ leads us to a reduction of the
$m_s$-value as determined from the data \cite{Chen} on the basis of
the earlier calculation \cite{Chetyrkin93}. The reduction by about
15\% leads to $m_s(M_\tau ) = (150 \pm 30_{exp} \pm 20_{th}$) MeV.  In
view of the large corrections the theoretical uncertainity can just be
considered as a guess based on Eq.~(\ref{num:disc:2.3}) and the
subsequent discussion.  This corresponds 
$m_s(\mbox{1 GeV} ) = 200 \pm 40_{exp} \pm 30_{th}$ MeV 
in good agreement with other determinations
(see, e.g.~\cite{others,ms:Jamin95,ms:as^2-analysis}).

\vspace{3mm}
\noindent
{\large \bf Acknowledgements}\\[2mm] This work is partially supported
by BMBF under Contract No. 057KA92P, DGF under Contract KU 502/8-1,
INTAS under Contract INTAS-93-744-ext and by Volkswagen Foundation
under Contract No.~I/73611. A.A.Pivovarov is supported in part by the
Russian Fund for Basic Research under contracts Nos.~96-01-01860 and
97-02-17065.

{\em Note added}.  Just before completing this paper we became aware
of a paper \cite{pich98} by Pich and Prades where mass corrections to
the total $\tau$ decay rate have been discussed.

\sloppy                                                                   
\raggedright                                                              
\def\app#1#2#3{{{\it} Act. Phys. Pol. }{{\bf} B #1} (#2) #3}                  
\def\apa#1#2#3{{{\it} Act. Phys. Austr.}{{\bf} #1} (#2) #3}                   
\def\lhc{Proc. LHC Workshop, CERN 90-10}                                    
\def\npb#1#2#3{{{\it} Nucl. Phys. }{{\bf} B #1} (#2) #3}                      
\def\plb#1#2#3{{{\it} Phys. Lett. }{{\bf} B #1} (#2) #3}                      
\def\partial#1#2#3{{{\it} Phys. Rev. }{{\bf} D #1} (#2) #3}                   
\def\pR#1#2#3{{{\it} Phys. Rev. }{ {\bf} #1} (#2) #3}                          
\def\prl#1#2#3{{{\it} Phys. Rev. Lett. }{{\bf} #1} (#2) #3}                   
\def\prc#1#2#3{{{\it} Phys. Reports }{{\bf} #1} (#2) #3}                      
\def\cpc#1#2#3{{{\it} Comp. Phys. Commun. }{{\bf} #1} (#2) #3}                
\def\nim#1#2#3{{{\it} Nucl. Inst. Meth. }{{\bf} #1} (#2) #3}                  
\def\pr#1#2#3{{{\it} Phys. Reports }{{\bf} #1} (#2) #3}                       
\def\sovnp#1#2#3{{{\it} Sov. J. Nucl. Phys. }{{\bf} #1} (#2) #3}              
\def\jl#1#2#3{{{\it} JETP Lett. }{{\bf} #1} (#2) #3}                          
\def\jet#1#2#3{{{\it} JETP Lett. }{{\bf} #1} (#2) #3}                         
\def\zpc#1#2#3{{{\it} Z. Phys. }{{\bf} C #1} (#2) #3}                         
\def\ptp#1#2#3{{{\it} Prog.~Theor.~Phys.~}{{\bf} #1} (#2) #3}                 
\def\nca#1#2#3{{{\it} Nouvo~Cim.~}{{\bf} #1A} (#2) #3}                        
\def\mpl#1#2#3{{{\it} Mod. Phys. Lett.~}{{\bf} A #1} (#2) #3}

\begin{table}
\caption{$m_s^2$ corrections to  $R_\tau$ and its moments; 
for the entries with $l=0$ the unknown coefficinet $k_3^{[2]}$ is set to zero}
\label{rsv:NAIVE}
\begin{center}
\begin{tabular}{|c|c|c|c|c|c|c|c|c|}
\hline
k & l & $b_0$ & $b_1$ & $b_2$ & $b_3$ & 
$\hat{b}_1$ & $\hat{b}_2$ & $\hat{b}_3$ 
\\
\hline
$0$ & $0$ & $-8$ & $5.333$ & $46.$ & $283.6$ & $3.034$ & $24.68$ & $92.82$
\\
$0$ & $1$ & $-\frac{5}{3}$ & $-4.167$ & $-112.8$ & $-1818.$ & $-6.466$ & $-112.3$ & $-1558.$
\\
$0$ & $2$ & $-\frac{3}{2}$ & $-1.767$ & $-54.9$ & $-711.4$ & $-4.066$ & $-59.89$ & $-605.8$
\\
$0$ & $3$ & $-\frac{7}{5}$ & $-1.1$ & $-40.24$ & $-474.3$ & $-3.4$ & $-46.76$ & $-408.5$
\\
$1$ & $0$ & $-\frac{75}{7}$ & $5.967$ & $56.59$ & $423.7$ & $3.667$ & $33.81$ & $202.9$
\\
$1$ & $1$ & $-\frac{9}{5}$ & $-5.767$ & $-151.4$ & $-2556.$ & $-8.066$ & $-147.2$ & $-2193.$
\\
$1$ & $2$ & $-\frac{21}{13}$ & $-2.433$ & $-69.57$ & $-948.5$ & $-4.733$ & $-73.02$ & $-803.2$
\\
$1$ & $3$ & $-\frac{3}{2}$ & $-1.481$ & $-48.58$ & $-601.7$ & $-3.781$ & $-54.22$ & $-513.2$
\\
$2$ & $0$ & $-\frac{27}{2}$ & $6.456$ & $65.25$ & $547.8$ & $4.156$ & $41.36$ & $302.7$
\\
$2$ & $1$ & $-\frac{21}{11}$ & $-7.433$ & $-192.3$ & $-3360.$ & $-9.733$ & $-184.3$ & $-2887.$
\\
$2$ & $2$ & $-\frac{12}{7}$ & $-3.148$ & $-85.31$ & $-1209.$ & $-5.447$ & $-87.12$ & $-1021.$
\\
$2$ & $3$ & $-\frac{27}{17}$ & $-1.898$ & $-57.57$ & $-740.8$ & $-4.197$ & $-62.26$ & $-627.9$
\\
$3$ & $0$ & $-\frac{49}{3}$ & $6.852$ & $72.61$ & $659.5$ & $4.553$ & $47.8$ & $393.8$
\\
$3$ & $1$ & $-2$ & $-9.148$ & $-235.1$ & $-4221.$ & $-11.45$ & $-223.1$ & $-3634.$
\\
$3$ & $2$ & $-\frac{9}{5}$ & $-3.898$ & $-101.9$ & $-1489.$ & $-6.197$ & $-102.$ & $-1256.$
\\
$3$ & $3$ & $-\frac{5}{3}$ & $-2.342$ & $-67.12$ & $-890.7$ & $-4.642$ & $-70.78$ & $-751.9$
\\
\hline
\end{tabular}
\end{center}
\end{table}
\begin{table}
\caption{$m_s^2$ corrections to the spin zero part of $R_\tau$}
\label{rsv:naive:J0}
\begin{center}
\begin{tabular}{|c|c|c|c|c|c|c|c|c|}
\hline
k & l & $b_0$ & $b_1$ & $b_2$ & $b_3$ & 
$\hat{b}_1$ & $\hat{b}_2$ & $\hat{b}_3$ 
\\
\hline
$0$ & $1$ & $\frac{3}{2}$ & $9.333$ & $110.$ & $1323.$ & $7.034$ & $79.47$ & $948.4$
\\
$0$ & $2$ & $\frac{3}{8}$ & $7.833$ & $72.96$ & $622.1$ & $5.534$ & $45.89$ & $346.7$
\\
$0$ & $3$ & $\frac{3}{20}$ & $7.233$ & $60.19$ & $427.2$ & $4.934$ & $34.5$ & $186.6$
\\
$1$ & $1$ & $\frac{9}{8}$ & $9.833$ & $122.3$ & $1556.$ & $7.534$ & $90.67$ & $1149.$
\\
$1$ & $2$ & $\frac{9}{40}$ & $8.233$ & $81.47$ & $752.$ & $5.934$ & $53.49$ & $453.5$
\\
$1$ & $3$ & $\frac{3}{40}$ & $7.567$ & $66.81$ & $518.3$ & $5.267$ & $40.36$ & $259.5$
\\
$2$ & $1$ & $\frac{9}{10}$ & $10.23$ & $132.5$ & $1757.$ & $7.934$ & $99.96$ & $1323.$
\\
$2$ & $2$ & $\frac{3}{20}$ & $8.567$ & $88.8$ & $868.9$ & $6.267$ & $60.05$ & $550.5$
\\
$2$ & $3$ & $\frac{3}{70}$ & $7.852$ & $72.66$ & $602.3$ & $5.553$ & $45.55$ & $327.5$
\\
$3$ & $1$ & $\frac{3}{4}$ & $10.57$ & $141.3$ & $1935.$ & $8.267$ & $107.9$ & $1477.$
\\
$3$ & $2$ & $\frac{3}{28}$ & $8.852$ & $95.26$ & $975.6$ & $6.553$ & $65.85$ & $639.7$
\\
$3$ & $3$ & $\frac{3}{112}$ & $8.102$ & $77.91$ & $680.4$ & $5.803$ & $50.23$ & $391.3$
\\
\hline
\end{tabular}
\end{center}
\end{table}
\begin{table}
\caption{$m_s^2$ corrections to the spin one part of $R_\tau$}
\label{rsv:naive:J1}
\begin{center}
\begin{tabular}{|c|c|c|c|c|c|c|c|c|}
\hline
k & l & $b_0$ & $b_1$ & $b_2$ & $b_3$ & 
$\hat{b}_1$ & $\hat{b}_2$ & $\hat{b}_3$ 
\\
\hline
$0$ & $1$ & $-5$ & $4.833$ & $35.73$ & $275.6$ & $2.534$ & $15.56$ & $113.$
\\
$0$ & $2$ & $-\frac{27}{8}$ & $3.567$ & $16.13$ & $29.42$ & $1.267$ & $-1.121$ & $-76.63$
\\
$0$ & $3$ & $-\frac{14}{5}$ & $3.067$ & $9.973$ & $-23.6$ & $0.767$ & $-6.129$ & $-111.$
\\
$1$ & $1$ & $-\frac{63}{10}$ & $5.376$ & $44.13$ & $381.1$ & $3.076$ & $22.71$ & $194.3$
\\
$1$ & $2$ & $-\frac{105}{26}$ & $3.967$ & $21.06$ & $71.84$ & $1.667$ & $2.885$ & $-49.16$
\\
$1$ & $3$ & $-\frac{13}{4}$ & $3.391$ & $13.55$ & $1.377$ & $1.091$ & $-3.295$ & $-97.15$
\\
$2$ & $1$ & $-\frac{84}{11}$ & $5.817$ & $51.34$ & $477.7$ & $3.517$ & $28.91$ & $270.3$
\\
$2$ & $2$ & $-\frac{33}{7}$ & $4.307$ & $25.49$ & $113.5$ & $2.007$ & $6.537$ & $-20.81$
\\
$2$ & $3$ & $-\frac{63}{17}$ & $3.674$ & $16.85$ & $26.68$ & $1.374$ & $-0.6516$ & $-81.99$
\\
$3$ & $1$ & $-9$ & $6.186$ & $57.65$ & $566.7$ & $3.886$ & $34.38$ & $341.5$
\\
$3$ & $2$ & $-\frac{27}{5}$ & $4.602$ & $29.53$ & $154.$ & $2.303$ & $9.891$ & $7.746$
\\
$3$ & $3$ & $-\frac{25}{6}$ & $3.925$ & $19.9$ & $51.96$ & $1.625$ & $1.825$ & $-66.$
\\
\hline
\end{tabular}
\end{center}
\end{table}
\begin{table}
\caption{$m_s^2$ corrections to  $R_\tau$ and its moments; 
 $\alpha_s(M_\tau) = 0.334$; 
the unknown coefficient $k_3^{[2]}$ is set to zero.}
\label{spin=:alphas(Mtau)=0.334}
\begin{center}
\begin{tabular}{|c|c|c|c|c|c|}
\hline
(k,l) & $\tilde{b}_0$  & $\tilde{b}_1$ & $\tilde{b}_2$ & $\tilde{b}_3$  & $\tilde{\delta}_{us,2}/\delta_{us,2}$ 
\\
\hline
$(0,0)$ & $-11.51$ & $2.539$ & $21.48$ & $50.14$ & $0.9325$
\\
$(0,1)$ & $3.768$ & $6.499$ & $90.06$ & $75.69$ & $2.18$
\\
$(0,2)$ & $-1.568$ & $4.643$ & $95.09$ & $71.91$ & $-4.184$
\\
$(0,3)$ & $-0.5285$ & $-2.855$ & $-80.94$ & $-4.813$ & $0.5966$
\\
$(1,0)$ & $-18.06$ & $2.893$ & $27.61$ & $52.42$ & $1.019$
\\
$(1,1)$ & $8.036$ & $6.209$ & $90.84$ & $75.1$ & $2.821$
\\
$(1,2)$ & $-2.768$ & $6.295$ & $133.9$ & $88.81$ & $-4.756$
\\
$(1,3)$ & $-0.1476$ & $-33.51$ & $-843.2$ & $-289.1$ & $2.851$
\\
$(2,0)$ & $-26.22$ & $3.211$ & $33.66$ & $54.59$ & $1.126$
\\
$(2,1)$ & $14.42$ & $6.219$ & $95.72$ & $76.65$ & $3.568$
\\
$(2,2)$ & $-5.014$ & $7.299$ & $158.5$ & $98.35$ & $-6.156$
\\
$(2,3)$ & $0.6619$ & $18.48$ & $459.2$ & $197.5$ & $4.709$
\\
$(3,0)$ & $-36.15$ & $3.504$ & $39.71$ & $56.74$ & $1.252$
\\
$(3,1)$ & $23.49$ & $6.326$ & $102.$ & $78.82$ & $4.453$
\\
$(3,2)$ & $-8.873$ & $7.866$ & $173.8$ & $103.4$ & $-8.211$
\\
$(3,3)$ & $2.204$ & $11.94$ & $299.5$ & $137.6$ & $7.139$
\\
\hline
\end{tabular}
\end{center}
\end{table}
\begin{table}
\caption{$m_s^2$ corrections to  $R^{(0)}_\tau$ and its moments; 
 $\alpha_s(M_\tau) = 0.334$,   the unknown coefficient $k_3^{[2]}$ is set to zero.}
\label{spin=J0:alphas(Mtau)=0.334}
\begin{center}
\begin{tabular}{|c|c|c|c|c|c|}
\hline
(k,l) & $\tilde{b}_0$  & $\tilde{b}_1$ & $\tilde{b}_2$ & $\tilde{b}_3$  & $\tilde{\delta}_{us,2}/\delta_{us,2}$ 
\\
\hline
$(0,1)$ & $4.779$ & $3.506$ & $39.55$ & $90.6$ & $1.274$
\\
$(0,2)$ & $0.2785$ & $-3.148$ & $-107.5$ & $-230.4$ & $-0.1803$
\\
$(0,3)$ & $0.2105$ & $2.363$ & $28.74$ & $53.57$ & $0.777$
\\
$(1,1)$ & $4.5$ & $3.918$ & $48.65$ & $110.5$ & $1.585$
\\
$(1,2)$ & $0.06802$ & $-20.2$ & $-529.1$ & $-1109.$ & $-0.6913$
\\
$(1,3)$ & $0.1234$ & $3.124$ & $50.53$ & $92.52$ & $1.041$
\\
$(2,1)$ & $4.432$ & $4.288$ & $57.51$ & $129.2$ & $1.955$
\\
$(2,2)$ & $-0.05534$ & $31.79$ & $763.$ & $1569.$ & $-1.388$
\\
$(2,3)$ & $0.08678$ & $4.085$ & $78.8$ & $142.1$ & $1.495$
\\
$(3,1)$ & $4.488$ & $4.628$ & $66.21$ & $146.9$ & $2.392$
\\
$(3,2)$ & $-0.1421$ & $14.87$ & $345.2$ & $697.9$ & $-2.318$
\\
$(3,3)$ & $0.06992$ & $5.115$ & $109.9$ & $195.2$ & $2.215$
\\
\hline
\end{tabular}
\end{center}
\end{table}
\begin{table}
\caption{$m_s^2$ corrections to  $R^{(1)}_\tau$ and its moments; 
 $\alpha_s(M_\tau) = 0.334$.}
\label{spin=J1:alphas(Mtau)=0.334}
\begin{center}
\begin{tabular}{|c|c|c|c|c|c|}
\hline
(k,l) & $\tilde{b}_0$  & $\tilde{b}_1$ & $\tilde{b}_2$ & $\tilde{b}_3$  & $\tilde{\delta}_{us,2}/\delta_{us,2}$ 
\\
\hline
$(0,1)$ & $-6.852$ & $1.861$ & $11.77$ & $98.8$ & $0.8833$
\\
$(0,2)$ & $-2.961$ & $0.9785$ & $-0.1989$ & $-70.28$ & $0.5589$
\\
$(0,3)$ & $-2.493$ & $1.257$ & $5.49$ & $41.2$ & $0.7861$
\\
$(1,1)$ & $-9.965$ & $2.071$ & $14.62$ & $139.$ & $0.9712$
\\
$(1,2)$ & $-3.5$ & $0.7497$ & $-4.874$ & $-161.9$ & $0.412$
\\
$(1,3)$ & $-3.026$ & $1.337$ & $6.927$ & $73.89$ & $0.8044$
\\
$(2,1)$ & $-13.78$ & $2.269$ & $17.54$ & $184.1$ & $1.081$
\\
$(2,2)$ & $-3.907$ & $0.36$ & $-12.71$ & $-318.4$ & $0.2254$
\\
$(2,3)$ & $-3.626$ & $1.458$ & $9.37$ & $131.9$ & $0.861$
\\
$(3,1)$ & $-18.39$ & $2.458$ & $20.54$ & $233.9$ & $1.213$
\\
$(3,2)$ & $-4.098$ & $-0.3007$ & $-25.99$ & $-589.4$ & $-0.01286$
\\
$(3,3)$ & $-4.322$ & $1.635$ & $13.16$ & $224.6$ & $0.9691$
\\
\hline
\end{tabular}
\end{center}
\end{table}

\end{document}